\documentclass[twocolumn]{aastex62}
\usepackage{CJK}

\newcommand{\cm}{\text{ cm}}

\newcommand{\hi}{\text{H}}

\newcommand{\g}{\text{g}}

\newcommand\aastex{AAS\TeX}

\graphicspath{{./}{figures/}}

\shorttitle{Modeling a Polluted White Dwarf with Cloudy}
\shortauthors{Steele et al. 2020}
\begin{document}
\begin{CJK}{UTF8}{gbsn}

\title{A Characterization of the Circumstellar Gas around WD 1124-293 using Cloudy}

\correspondingauthor{Amy Steele}
\email{asteele@astro.umd.edu}

%\author[0000-0002-0786-7307]{Greg J. Schwarz}
%\affil{American Astronomical Society \\
%2000 Florida Ave., NW, Suite 300 \\
%Washington, DC 20009-1231, USA}

\author[0000-0002-0141-7946]{Amy Steele}
\affiliation{University of Maryland at College Park} %\\
%4296 Stadium Dr. \\
%College Park, MD 20742-2421, USA}
%\collaboration{(AAS Journals Data Scientists collaboration)}
%\nocollaboration

\author[0000-0002-1783-8817]{John Debes}
\affiliation{Space Telescope Science Institute}
%\nocollaboration

\author[0000-0002-8808-4282]{Siyi Xu (许\CJKfamily{bsmi}偲\CJKfamily{gbsn}艺)}
\affil{NSF's NOIRLab/Gemini Observatory, 670 N. A'ohoku Place, Hilo, Hawaii, 96720, USA }
%\affil{Gemini Observatory, 670 N. A'ohoku Place, Hilo, HI 96720}
%\affil{NSF's National Optical-Infrared Astronomy Research (NOIR) Laboratory}

\author[0000-0002-4037-3114]{Sherry Yeh}
\affiliation{W. M. Keck Observatory}

\author{Patrick Dufour}
\affiliation{Universit\'{e} de Montreal}

%% Note that the \and command from previous versions of AASTeX is now
%% depreciated in this version as it is no longer necessary. AASTeX 
%% automatically takes care of all commas and "and"s between authors names.

%% AASTeX 6.2 has the new \collaboration and \nocollaboration commands to
%% provide the collaboration status of a group of authors. These commands 
%% can be used either before or after the list of corresponding authors. The
%% argument for \collaboration is the collaboration identifier. Authors are
%% encouraged to surround collaboration identifiers with ()s. The 
%% \nocollaboration command takes no argument and exists to indicate that
%% the nearby authors are not part of surrounding collaborations.

%% Mark off the abstract in the ``abstract'' environment. 
\begin{abstract}
Between 30 - 50\% of white dwarfs (WDs) show heavy elements in their atmospheres. This ``pollution" is thought to arise from the accretion of planetesimals perturbed by outer planet(s) to within the WD's tidal disruption radius. A small fraction of these WDs show either emission or absorption from circumstellar (C-S) gas. The abundances of metals in the photospheres of WDs with C-S gas are mostly similar to the bulk composition of the Earth. The C-S component arises from gas produced through collisions and/or the sublimation of disintegrating planetesimals. High resolution spectroscopic observations of WD 1124-293 reveal photospheric and C-S absorption of Ca in multiple transitions. Here, we present high signal-to-noise ratio spectra, an updated WD atmosphere analysis, and a self-consistent model of its C-S gas. We constrain the abundances of Ca, Mg, and Fe in the photosphere of WD 1124-293, and find agreement with the abundances of these 3 species in the C-S gas. We find the location of the C-S gas is about a hundred white dwarf radii, the C-S and photospheric compositions are thus far consistent, the gas is not isothermal, and the amount of C-S Ca has not changed in two decades. We also demonstrate how to use Cloudy to model C-S gas viewed in absorption around polluted WDs. Modeling the abundances of C-S gas around polluted WDs with Cloudy provides a new method to measure the composition of exo-planetesimals and will allow a direct comparison to the composition of rocky bodies in the Solar System.

\end{abstract}

%% Keywords should appear after the \end{abstract} command. 
%% See the online documentation for the full list of available subject
%% keywords and the rules for their use.

%% From the front matter, we move on to the body of the paper.
%% Sections are demarcated by \section and \subsection, respectively.
%% Observe the use of the LaTeX \label
%% command after the \subsection to give a symbolic KEY to the
%% subsection for cross-referencing in a \ref command.
%% You can use LaTeX's \ref and \label commands to keep track of
%% cross-references to sections, equations, tables, and figures.
%% That way, if you change the order of any elements, LaTeX will
%% automatically renumber them.
%%
%% We recommend that authors also use the natbib \citep
%% and \citet commands to identify citations.  The citations are
%% tied to the reference list via symbolic KEYs. The KEY corresponds
%% to the KEY in the \bibitem in the reference list below. 

\section{Introduction} \label{sec:intro}

The spectra of white dwarfs (WDs) should show only pressure-broadened hydrogen and/or helium absorption lines, yet at least 27\% of young WDs with temperatures less than $\sim$27,000 K have photospheres polluted by elements heavier than helium \citep{Koester2014}. These metals should settle out of the atmospheres of WDs on timescales of days to Myr depending on the WD temperature, surface gravity, and main atmospheric composition (H vs. He) \citep{Koester2009}. For isolated WDs, the pollution could arise from grains in the interstellar medium (ISM) (\citealt{Dupuis1993a,Dupuis1993b}), or more likely, from the accretion of solids that have been liberated from a captured planetesimal (e.g. WD 1145+017l \citealt{Vanderburg2015}). The gas phase of the latter accretion process can exist as a circumstellar disk observable as a double peaked line in emission (e.g. \citealt{Manser2016}) or a Doppler broadened profile in absorption (e.g. \citealt{Xu2016}). WD 1124-293 is one of a few WDs that shows both metal photospheric absorption and circumstellar absorption features. %To date, there are only two polluted WDs that show both circumstellar and photospheric absorption at optical wavelengths: WD1145+017 and WD 1124-293 \citep{Debes2012} [unclear about Boris's star...]. Accretion from the ISM has been ruled out in both cases. 

%\subsection{The Polluted Photosphere Model for WD 1124-293}
%Something here that makes sense.... 

%Patricks text (slightly edited by Amy):

We investigate the gas toward WD 1124-293 to explore the conditions necessary to produce the observed absorption. WD 1124-293 is of spectral type DAZ (a hydroden dominated atmosphere with metal lines), has an effective temperature $T_{\text{eff}} \sim 9367$K, and surface gravity $\log g = 7.99$ (see Table \ref{tab:properties}). The one observed C-S Ca K absorption feature detected at $8\sigma$ \citep{Debes2012} provides an opportunity to explore the physical conditions that result in this type of spectrum. 
We use the microphysics code, Cloudy, to model the metal-rich gas polluting WD 1124-293 by creating a grid of models of C-S gas to explore the abundances of elements from He to Zn relative to hydrogen. Due to the lack of an infrared excess \citep{Barber2016}, we exclude grains from our models. From the code, we obtain line optical depths, species column densities, and the temperature profile through the gas cloud. With these models, we place constraints on the potential masses and abundances of the C-S gas. %Siyi edit: that could result in a spectrum dominated by calcium species for WD 1124-293.

%We show how the gas polluting WD 1124 can be modeled using the microphysics code, Cloudy.  We created a Cloudy grid of models forC-S gas around WD 1124 to explore the abundances of elements from He to Zn relative to hydrogen, and obtain line optical depths, species column densities, and the temperature profile through the gas cloud. With these models, we place constraints on the potential masses and abundances that could result in a spectrum dominated by calcium species for WD 1124-293.

In Section 2, we present new observations of WD 1124-293 with Keck HIRES and an improved co-added MIKE spectrum. In Section 3, we describe how we know the pollution of WD 1124-293 visible in the new, higher resolution HIRES spectrum is not due to the ISM. In section 4, we describe how we model a polluted white dwarf with Cloudy and apply this method to WD 1124-293, showing how we can determine the characteristics of its C-S gas using Cloudy. In short, we build a grid of models and place constraints on the column densities needed for detecting features. In section 5, we present our results and conclude in section 6.

\begin{deluxetable}{lcC}%[h!]
\tablecaption{Properties of WD 1124-293 \label{tab:properties}}
\tablecolumns{3}
\tablenum{1}
\tablewidth{0pt}
\tablehead{
\colhead{Parameter } & \colhead{Value} & \colhead{Reference} 
}
\startdata
RA, Dec (J2000) & 11:27:09.25, $-29$:40:11.20 &  \tablenotemark{a} \\
Spectral Type & DAZ &  \tablenotemark{b} \\
$G$ (mag) & 15.0139 & \tablenotemark{a} \\
$T_{\text{wd}}$ (K) & 9367 &  \tablenotemark{c} \\
$\log g$ (cm s$^{-2}$)   & 7.99  &  \tablenotemark{c}   \\
$R_{\text{wd}}$ ($R_\odot$) & 0.0127  & \tablenotemark{c}  \\
$M_{\text{wd}}$ ($M_\odot$) & 0.575  & \tablenotemark{c}  \\
Luminosity ($L_\odot$) & 0.00111  & \tablenotemark{c}    \\
Distance (pc) &  33.691  & \tablenotemark{a}   \\
\enddata
\tablecomments{$^{a}$\cite{GaiaDR2}, $^{b}$\cite{Subasavage2017}, $^{c}$from \url{http://www.montrealwhitedwarfdatabase.org/evolution.html}}
\end{deluxetable}

%%%%%%%%%%%%%%%%%%%%
\section{Observations} \label{sec:style}

% Given this spectral resolution, we can resolve the photospheric component from the circumstellar component, if present. 

WD 1124-293 was observed 16 times between 1998 and 2011 \citep{Debes2012} with the MIKE echelle spectrograph \citep{Bernstein2003}. We observed WD 1124-293 using the HIRES echelle spectrograph \citep{Vogt1994} on the Keck I telescope for 1200s on April 24, 2018. For  the HIRES observations, the blue collimator and C5 decker were used with a slit width of 1.148 arcsec, typically giving a spectral resolution of $\sim\!40,000$. However, the seeing during the observation was $\sim0.5$ arcsec, and the effective resolution is $\sim\!80,000$. The data were reduced with the MAKEE package. The Ca K line was extracted over two orders and averaged, while the H line was extracted over one order. We continuum-normalized 10~\AA{} regions of the spectrum by fitting a polynomial to the continuum to remove effects due to the instrument response function. The signal-to-noise (S/N) ratio is $\sim38$ in a 10~\AA{} region (3935~\AA{} to 3945~\AA{}), which corresponds to $\sim75$ per resolution element. We detect C-S and photospheric absorption at the Ca K line and only photospheric absorption at the Ca H line. We report that no new absorption or emission features have been detected.

%(Amy: SNR was caluclated by taking the spectrum from 3920 - 3930 and calculating the standard deviation of points. A resolution element is $\sim$2 pixels so $194=\SQRT(2)*137$). 

\begin{deluxetable}{cccc}%[b!]
\tablecaption{Keck/HIRES Observations of WD 1124-293 and Standards \label{tab:observations}}
\tablecolumns{4}
\tablenum{3}
\tablewidth{0pt}
\tablehead{
 \colhead{Target} & \colhead{Dist (pc)} &  \colhead{Sep ('')} & \colhead{Exp (s)}
}
\startdata
WD 1124-293  & 33.7 & -- &1200s   \\
HIP56280A   &   26.3  & 79  & 30s   \\
HIP56280B   &   26.3  &  79  & 30s   \\
HIP55864    &  117   & 16  &100s  \\
HIP55901    &  401  &  24.6 & 100s  \\
\enddata
%\tablenotetext{a}{At exposure start.}
\tablecomments{The distance, separation from WD 1124-293, and exposure time for the target and standards. The observation date was April 24, 2018, at a resolution $\sim\!40,000$ with a wavelength range of 3100 - 5950 \AA.}
\end{deluxetable}

Additionally, we reanalyzed the many epochs of spectra presented in \cite{Debes2012}. The multiple MIKE epoch spectroscopy was extracted over two orders that contained the Ca H and K lines. For this work, we took the reduced spectra of WD 1124-293 from both the red and blue CCDs at each epoch and flux calibrated the orders against a model DA atmosphere \citep{Koester2010}\footnote{\url{http://svo2.cab.inta-csic.es/theory/newov2/index.php}} with the appropriate $T_{\text{eff}}$ and $\log g$ (see Table \ref{tab:properties}) based on WD 1124-293's parallax and spectro-photometry from APASS, 2MASS, and ALLWISE catalogs. Once all epochs were flux calibrated, we median combined the spectra to a common wavelength grid with a sampling of 0.05 \AA{} between 3350~\AA{} and 5030~\AA{} and 0.08 \AA{} between 5030~\AA{} and 9400~\AA{}. The typical resolution for the MIKE spectra is $\sim34,000$. Near the Ca K line, the final S/N ratio of the data as measured in the continuum corresponds to $\sim$194 per resolution element.

\begin{figure*}
    \centering
      \includegraphics[scale=0.6]{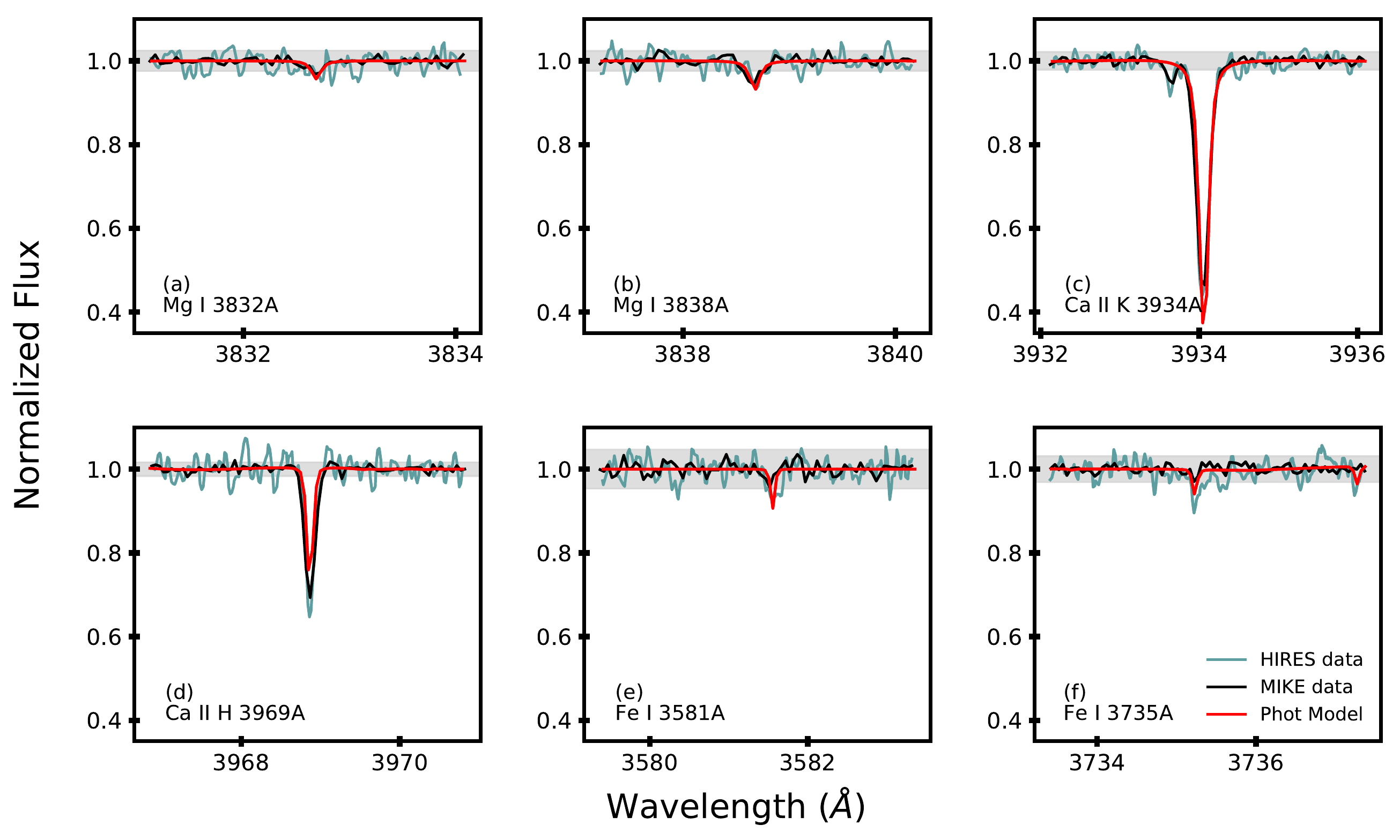}
    \caption{Detection of metal photospheric absorption lines for WD 1124-293 using a co-added MIKE spectrum. For each square panel, the MIKE data are shown in black, new Keck HIRES data are shown in teal, and a model of the photosphere is shown in red. The model is convolved with a Gaussian kernel with a width determined by a resolution, $R=60,000$ using a routine from PyAstronomy (\url{https://github.com/sczesla/PyAstronomy}, \citealt{Czesla2019}). The gray region highlights a $\pm3\sigma$ region for the MIKE spectra.}
    \label{fig:remake_dufour}
\end{figure*}

\begin{deluxetable*}{lccccccccccc}
\tablecaption{Ca absorption line properties for WD 1124-293 \label{tab:ca_properties}}
\tablecolumns{12}
\tablenum{2}
\tabletypesize{\scriptsize} %footnotesize}
\tablewidth{300pt}
\tablehead{
\\
&\multicolumn{11}{c}{Ca K : $\lambda_0 = 3933.6614$~\AA{}, Ca H : $\lambda_0 = 3968.4673$~\AA{}}  \\
 &\multicolumn{4}{c}{Circumstellar} & & \multicolumn{6}{c}{Photospheric} \\
 \cline{2-5}
 \cline{7-12}
\colhead{Dataset} & \colhead{$\lambda_c$} & \colhead{$v_c$} & \colhead{Eq Width} & \colhead{FWHM} & & \colhead{$\lambda_c$} & \colhead{$v_c$} & \colhead{Eq Width} & \colhead{$\lambda_c$} & \colhead{$v_c$} & \colhead{Eq Width} \\
 & \colhead{(\AA)} & \colhead{(km/s) } & \colhead{(m\AA)} & \colhead{(km/s)} & & \colhead{(\AA)} & \colhead{(km/s) } & \colhead{(m\AA) } & \colhead{(\AA)} & \colhead{(km/s) } & \colhead{(m\AA)}  
}
\startdata
\\ %MIKEa & 3933.649 	&  $-1.1\pm0.6$ 
\hline
%MIKEa & 3933.649 	&  $-1.1$ &  11		& $15.2$ & & 3934.043 & 29.1 & $113\pm0.6$ & 3934.0262 & 27.8 & $52\pm0.6$ \\
MIKE & $3933.623$ & $-2.9$ &7.1 	& $9.91$ & & 3934.020 & 27.3 & $113\pm2$ & 3968.841 & 28.25 & $52\pm1$\\
HIRES	& $3933.637$ & $-1.9$ &7.8  	& $5.97$ & & 3934.053 & 29.8 & $105\pm1.5$ & 3968.879 & 31.1 & $65\pm 3$ 
\enddata
\tablecomments{The Ca H and K line wavelengths are given in air. We only detect a C-S feature at the Ca K line. For the C-S line, the central wavelength of the line is $\lambda_c$, the Doppler shift velocity is $v_c$, the equivalent width of the line assuming a Gaussian profile is ``Eq Width," and the full width at half max is FWHM.}
\end{deluxetable*}

%We present the Ca absorption line properties for WD 1124-293 in Table \ref{tab:ca_properties} for the spectra presented in \citealt{Debes2012} (referred to as MIKEa), and the co-added MIKE spectrum (MIKEb), and Keck HIRES spectrum presented in this work..... The MIKEa properties in Table \ref{tab:ca_properties} are averages of the individual spectra versus the properties of the one co-added MIKEb spectrum. 
We present the Ca absorption line properties for WD 1124-293 in Table \ref{tab:ca_properties} for the co-added MIKE spectrum and Keck HIRES spectrum presented in this work. We fit the circumstellar and photospheric absorption features with Gaussians to calculate the equivalent widths of the lines, as well as the FWHMs. The photospheric lines are gravitationally red-shifted. All line center velocities for the Ca K and H lines are also presented in Table \ref{tab:ca_properties}. The HIRES dataset has the highest resolution of the three sets, and so we use it to constrain the location of the gas. The co-added MIKE spectrum has the highest S/N, so we use it to constrain the abundances of different species. In Figure \ref{fig:remake_dufour}, we show selected photospheric absorption features due to Mg, Ca, and Fe that appear in both the co-added MIKE spectrum and new HIRES spectrum.

%%%%%%%%%%%%%%%%%%%%%%%%%%%
\begin{figure}
    \centering
      \includegraphics[scale=0.5]{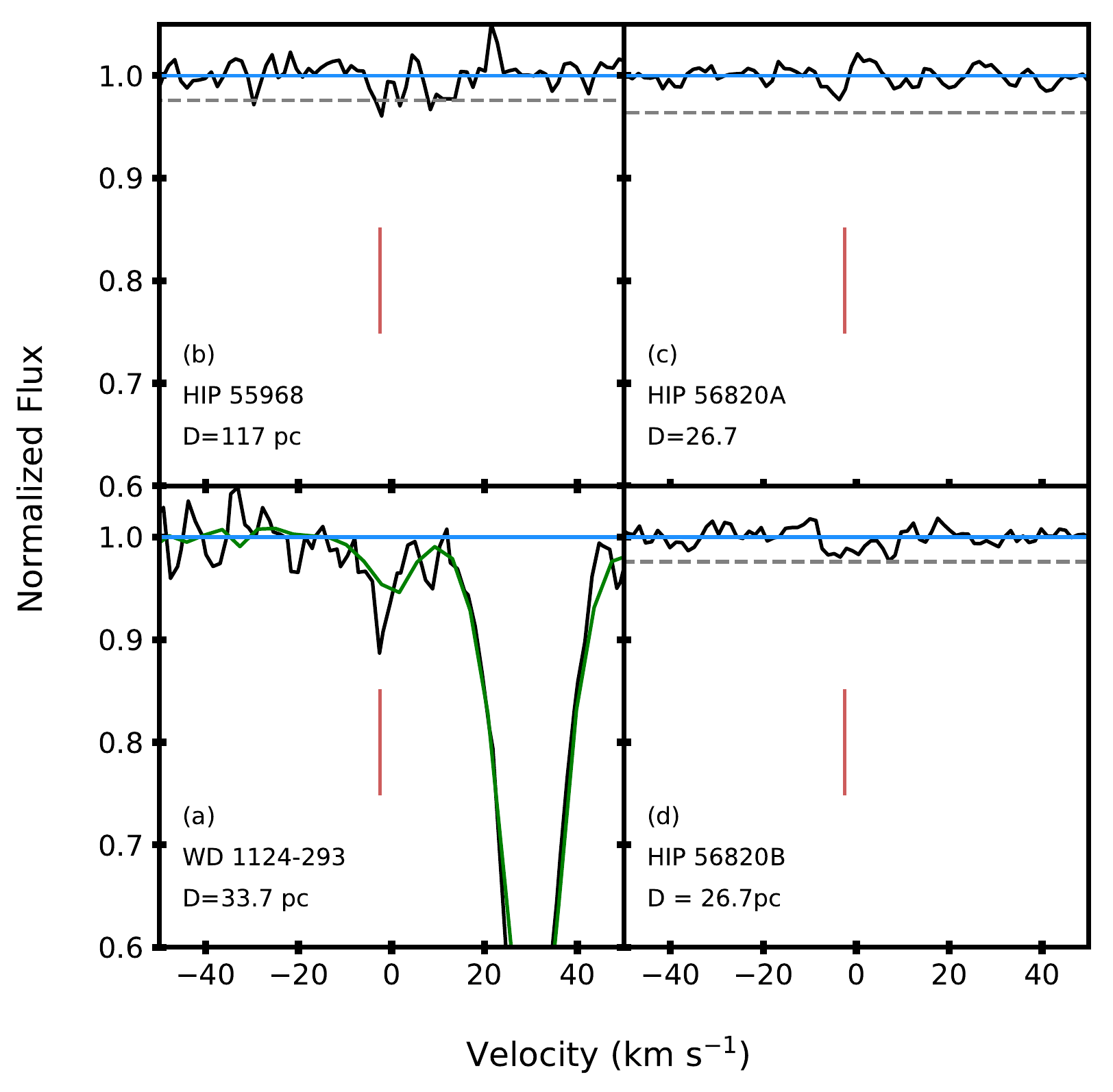}
    \caption{The normalized flux for the (a) target and nearby sources (b) - (d) HIP 56280A, HIP 56280B, and HIP 55864. For all panels, the blue line is drawn at a value of 1.0 to guide the eye, the 2019 HIRES data are shown in black, and the red vertical line gives the approximate velocity of the C-S line for WD 1124-293, -2.5 km s$^{-1}$. In panels (b) - (d), the dashed gray is drawn at $3\sigma$ below the continuum. In panel (a), the green curve is the normalized flux from the Magellan MIKE observations described in \cite{Debes2012}.}
    %There is a slight shift in the central wavelength of the between the two datasets that we discuss in section~\ref{sec:disc} .}
    \label{fig:remake}
\end{figure}

%%%%%%%%%%%%%%%%%%%%%%%%%%%
%%%%%%%%%%%%%%%%%%%%%%%%%%%

\section{ Pollution and the ISM}
\cite{Debes2012} first investigated the possibility that the weak Ca absorption detected towards WD 1124-293 could be explained by coincident local ISM absorption by comparing against high SNR spectra of stars located closely in the sky. For that study, \cite{Debes2012} looked at HIP 56280A, HIP 55864, HIP 55731, HIP 55901, and HIP 55968, but did not have sufficient parallax information on all of the stars to correctly sort them in terms of increasing distance from the Earth. With the advent of the Gaia mission, secure parallaxes now exist for that sample of stars. We re-observed HIP 56280A, HIP 55901, and HIP 55864 and in addition observed HIP 56280B, the physically bound companion to HIP 56280A. HIRES has higher spectral resolution compared to MIKE, and thus higher sensitivity to weak spectral features. The addition of HIP 56280B also ensures tighter constraints on the amount of Ca present in the ISM interior to $\sim\!30$ pc with two independent measurements of that part of the sky.

In order to search for weak lines from the ISM we first had to fit and remove the continuum near the expected rest frame velocity of the Ca K line in question for each standard star. This was relatively straightforward for HIP 55901 and HIP 55864, which show broad Ca absorption due to the rapid rotation of the host stars--the continuum can be fit with a high order polynomial \citep{Debes2012}.

Our approach for HIP 56820A and HIP56820B was slightly different, due to the later spectral type of these two stars. Both objects are roughly consistent with F stars and are likely nearly the same effective temperature and gravity. Both stars show Ca emission in the line core due to stellar activity, though HIP56820A shows stronger emission. For both stars we fit the broad Ca component with a spline fit and then fit the Ca line core with a two component Gaussian curve. We verified that our fits did not unintentionally fit any absorption lines coincident with the rest velocity of the C-S  line seen in WD 1124-293. Our resulting continuum fit lines are shown in Figure ~\ref{fig:remake}, along with the expected $3\sigma$ upper limit to detectable absorption for each star. We estimated this upper limit by taking the standard deviation of flux in the normalized spectra, assuming that we would detect anything 3$\sigma$ below the continuum level. We note that we do not show the spectrum of HIP 55901, which shows strong absorption consistent with that observed previously in \cite{Debes2012}. None of these comparison stars near WD 1124-193 show Ca absorption at the velocity of the observed Ca line, so we confidently rule out an ISM contribution to the WD spectrum.

% The previous last sentence below is confusing to me, so I swapped it with the current final sentence of the above section:
%In all cases a Ca line with a depth consistent with that seen around WD 1124-293 is at least twice as deep as this upper limit, thus ruling out an ISM contribution to the observed line. 

%%%%%%%%%%%%%%%%%%%%%%%%%%%
%%%%%%%%%%%%%%%%%%%%%%%%%%%
\section{Modeling WD Circumstellar Gas with Cloudy} \label{sec:floats}
%, allowing for a comparison to the compositions of other bodies (e.g., the Earth, chondrites, comets, etc.)
 %Calculations were performed with version 17.01 of Cloudy, last described by Ferland et al. (2017).

%We can also compare the abundances of metals in the C-S gas to the abundances of metals already accreted in the photosphere. 

%In this section, we describe how we determine the abundances of metals in the photosphere and for the C-S gas, where we introduce our methods for applying Cloudy to polluted WD modeling. 

\subsection{Cloudy Model Inputs}
We aim to constrain the abundances of metals in the circumstellar gas of WD 1124-293. A Cloudy input file requires an ionizing source, the geometry of the gas, the density of hydrogen in the gas, and the abundances of He to Zn, relative to H. In this section, we describe how we determine these inputs for WD 1124-293 and how we use the Cloudy output, thereby describing our method for using Cloudy to model the C-S environment of a polluted white dwarf. Calculations were performed with version 17.01 of Cloudy, last described by \cite{Ferland2017}.
%Something here that makes sense.... 

%Patricks text (slightly edited by Amy): (moved to results)
%To obtain the abundances of heavy elements, we proceed in a similar way as described in Xu et al. 2019, by computing grids of synthetic spectra for each element of interest using a pure DA atmospheric structure assuming the stellar parameters previously determined by fitting the photometric data and parallax measurement ($T_{\text{eff}} = 9367$ K and $\log g = 7.99$). The abundances are then obtained by minimizing $\chi^2$ between the normalized spectroscopic data and the grid spectrum. Trace amounts of metals have a negligible effect on the thermodynamic structure for the grid spectrum. 

\subsubsection{The ionizing source}
 For the Cloudy model, we use an interpolated Koester DA photosphere \citep{Koester2010,Tremblay2009}\footnote{Models were downloaded from the Spanish Virtual Observatory database.} with a temperature of 9420 K and luminosity of $0.00111 L_\odot$ as the input, or ionizing continuum.  The continuum is mapped to an energy mesh with a resolution of 0.05, (20x the native course grid of $R\sim300$). % Siyi edit: We next discuss the geometry of the gas, and then the composition.
 
% We also choose to set the outer edge of the gas disk to the tidal radius for a polluted white dwarf, which for WD 1123-293 is $R_{\text{tide}}\sim100\,R_{\text{wd}}$. This radius is an upper limit or stopping criterion for all Cloudy model calculations.

%The outer edge of the gas could be set by the sublimation radius or a pile-up radius, $r_{\text{pile}}$,
%\begin{equation}
%    r_{\text{pile}} = 6.89\times10^{11}(1 + x^{-1})^{1/2} \left(\frac{L_\star}{L_\odot}\right)^{1/2}\left(\frac{1300\ke}{T_{\text{sub}}}\right)^2
%\end{equation}
%where $x$ is a constant that depends on the composition of the material sublimating %\citep{Kobayashi2011}. 

\subsubsection{Geometry}
The geometry of the C-S gas around WD 1124-293 is unknown. We choose to set the geometry in a disk, rather than a sphere because we have shown that it is highly unlikely that the ISM is the source of the pollution. \cite{Debes2012} place constraints on the location of the C-S  gas, with a minimum distance of $7^{+11}_{-3}\,R_{\text{wd}}$, maximum distance of 32000 AU, and dynamical estimate of $\sim$54 $R_{\text{wd}}$, where $R_{\text{wd}}$ is the radius of WD 1124-293. The dynamical estimate of the location of the gas is determined assuming the gas is in a circular Keplerian orbit and using the FWHM of the gas absorption feature to determine the upper limit to the disk's orbital velocity. The ``dynamical" distance of the gas from the WD, $R_{\text{Kep}}$, is 
\begin{equation}
    R_{\text{Kep}} \approx (GM_{\text{wd}})^{1/3}\left[\frac{2R_{\text{wd}}}{\text{FWHM (cm/s)}}\right]^{2/3},
\end{equation}
where $G$ is the gravitational constant. From the new HIRES dataset, we measure the FWHM of the C-S absorption feature assuming the line profile is Gaussian. With the resulting FWHM $\sim6$ km/s, and the radius listed in Table \ref{tab:properties}, $R_{\text{Kep}}\sim 106\, R_{\text{wd}}$ for WD 1124-293, a value almost double that in \cite{Debes2012} due to the high spectral resolution of the HIRES data. $R_{\text{Kep}}$ is the best estimate of the minimum distance of the gas to the star and it depends on the width of the absorption feature. If the C-S absorption feature were more narrow, then $R_{\text{Kep}}$ would be larger. %$\log (R_{\text{wd}} \cm^{-1}) = 8.92$

\begin{figure}
    \centering
    \includegraphics[scale=0.33]{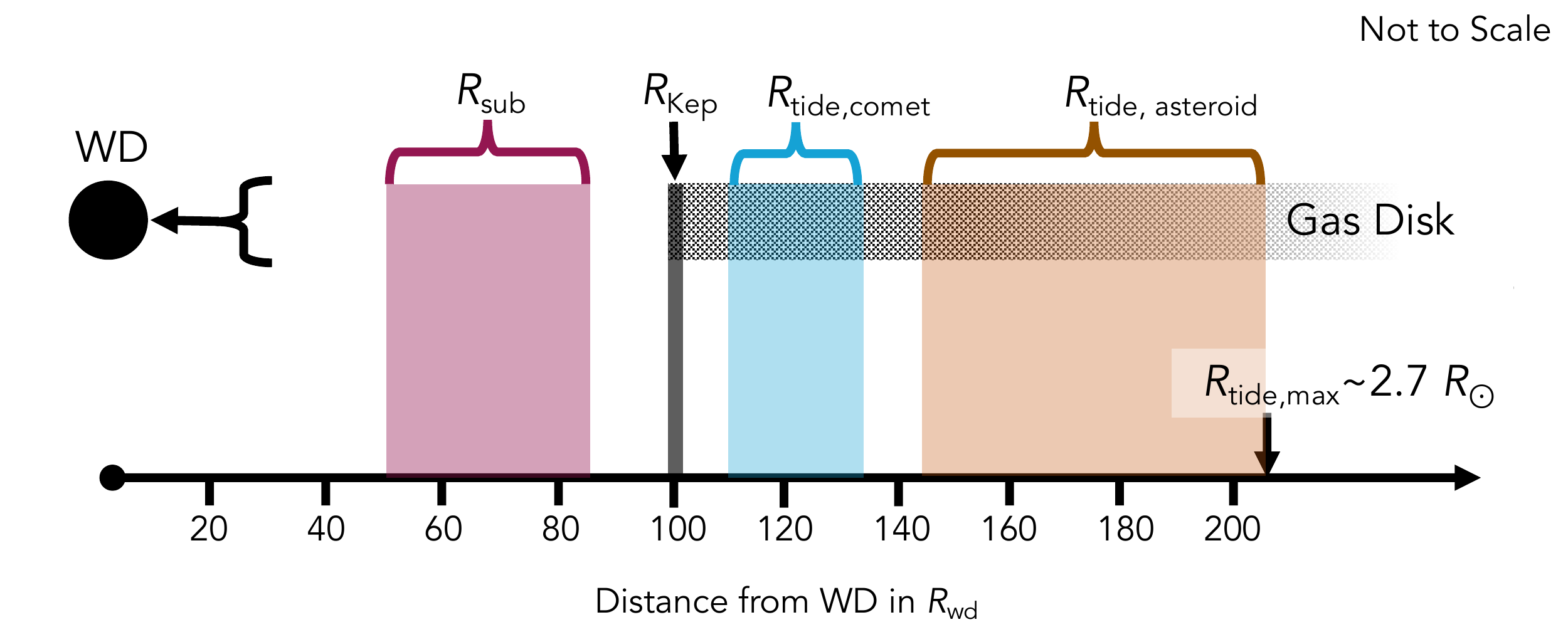}
    \caption{A not-to-scale cartoon of the model for WD gas and debris disks inspired by \cite{Metzger2012}, where a solid debris disc typically forms near the tidal disruption radii which happens to be $\sim R_\odot \sim 120 R_{\text{wd}}$. There is no evidence that WD 1124-293 has a dusty debris disk, so we exclude it. The gas is shown with the speckled black and white region. We show 4 different radii $R_{\text{sub}}, R_{\text{Kep}}, R_{\text{tide,comet}}, R_{\text{tide,asteroid}}$ and a range of their possible values, with the shaded regions brought down to the axis for ease of reading, where applicable. The outer extent of the gas is unknown, as is the exact accretion mechanism. The maximum tidal radius for WD 1124-293 is $\sim211R_{\text{wd}}\sim 2.7R_\odot$.  
    }
    \label{fig:metzger}
\end{figure}

The outer edge of the gas is unknown, but we can consider a sublimation radius and a tidal disruption radius. The sublimation radius, $R_{\text{sub}}$ is the distance at which the equilibrium temperature of particles equals their sublimation temperature, $T_{\text{sub}}$, with
\begin{equation}
    R_{\text{sub}} = \frac{R_{\text{wd}}}{2}\left(\frac{T_{\text{wd}}}{T_{\text{sub}}}\right)^2.
\end{equation}
This equation represents the smallest possible sublimation radius for optically thin distributions of dust and it assumes the particles absorb and emit radiation perfectly. $R_{\text{sub}}$ depends on the shape, size, and composition of the particle. Graphite grains (0.01$\mu$m) and astronomical silicate grains (0.1 $\mu$m) would sublimate at 54 $R_{\text{wd}}$ and 93 $R_{\text{wd}}$, respectively (see Figure~\ref{fig:metzger}). In Section~\ref{sec:disc}, we discuss why the detectable gas at $R_{\text{Kep}}$ is farther than $R_{\text{sub}}$.

The tidal disruption radius, $R_{\text{tide}}$ \citep{Davidsson1999,Jura2003,Veras2014}, also depends on the composition of the material with
\begin{equation}
    \frac{R_{\text{tide}}}{R_\odot} = C_{\text{tide}} \left(\frac{M_{\text{wd}}}{0.6 M_\odot}\right)^{1/3} \left(\frac{\rho_b}{3 \text{ g cm}^{-3}}\right)^{-1/3} %\left(\frac{\rho_{\text{wd}}}{\rho_b}\right)^{1/3}R_\star
\end{equation}
where $C_{\text{tide}}$ has typical values of 0.85 to 1.89 \citep{BearSoker2013}, and $\rho_b$, the density of the disrupting body satisfies $\rho_b \geq 1 \,\g\cm^{-3}$ \citep{Carry2012,Veras2014}. For WD 1124-293, the maximum tidal disruption radius is $R_{\text{tide}}\sim\!200\, R_{\text{wd}}$. We show typical ranges of tidal disruption radii for comets and asteroids ($R_{\text{tide,comet}}$ and $R_{\text{tide,asteroid}}$) in Figure~\ref{fig:metzger}.\footnote{For comparison, for comet Shoemaker Levy 9 had $C_{\text{tide}} \sim 1.31$ and $\rho_{b}\lesssim0.702$ \citep{Boss1994}, and would have a disruption radius of $\sim110\,R_{\text{wd}}$ near WD 1124-293.}

%where $C_{\text{tide}}$ has typical values of 0.85 to 1.89 \citep{BearSoker2013}, $\rho_b$, the density of the disrupting body satisfies $\rho_b \geq 1 \,\g\cm^{-3}$ \citep{Carry2012,Veras2014}, and $\rho_{\text{wd}}$, the density of WD 1124-293 is $\sim 3.96\times10^5 \,\g\cm^{-3}$, resulting in a maximum tidal disruption radius of $r_{\text{tide}}\sim\!200\, r_{\text{wd}}$.\footnote{For comparison, for comet Shoemaker Levy 9 had $C_{\text{tide}} \sim 1.31$ and $\rho_{b}\lesssim0.702$ \citep{Boss1994}, and would have a disruption radius of $\sim110\,r_{\text{wd}}$ near WD 1124-293.}

%Move to discussion? Yes. The detectable gas at $r_{\text{Kep}}$ is farther than $r_{\text{sub}}$. Gaseous disks can accrete inward and spread outward due to angular momentum transport by turbulent viscosity \citep{Metzger2012}. 

We take $R_{\text{Kep}}$ as our minimum radius of the gas and assume that these radii follow the relation $R_{\text{tide}} > R_{\text{Kep}} > R_{\text{sub}}$ (see Figure~\ref{fig:metzger}). We set the gas to extend from $100\,R_{\text{wd}}$ to $200\,R_{\text{wd}}$ (approximately $R_{\text{Kep}}$ to $R_{\text{tide}}$). The aspect ratio of the gas disk is $h/r\sim10^{-3}$ \citep{Metzger2012}, so  we truncate the gas to a cylinder with a height of 10\% $R_{\text{wd}}$. We discuss the implications of this relation among radii in Section~\ref{sec:disc}.  

%There are two patterns to highlight that we have no observational evidence for gas inward of $R_{\text{Kep}}$.

%Tidally disrupted solids initially migrate inward due to Poynting-Robertson drag and sublimate, forming gas. The gas then accretes inward and spreads outward.

%Cloudy uses the number density of hydrogen to set the conditions of a cloud, so we set our abundances relative to hydrogen. Due to the lack of an infrared excess, we exclude grains.

%Cloudy uses the number density of hydrogen to set the conditions of a cloud, so abundances in the models are set relative to hydrogen. Due to the lack of an infrared excess, we exclude grains.

%In a given Cloudy model, we can individually set the abundances of H through Zn. 

%set such that $\log n(El)/n(\text{H})$ is equal to the value it would have if the abundance of $El$ were its solar [CHECK ME].

\subsubsection{Hydrogen density}

Most planetesimals accreted onto WDs are water poor, so there should be very little hydrogen gas present around WD 1124-293, assuming a planetesimal origin (see section 5.3 of \citealt{JuraYoung2014}, and references therein). However, Cloudy uses the number density of hydrogen, $n_{\text{H}}$, to set the conditions of a cloud. If a C-S gas spectrum has features due to hydrogen, one could probe the maximum amount of hydrogen present in the system by calculating the column density due to H, $N_{\text{H}}$,  assuming every H atom participates in the line transition. With no such features and a gas temperature too low for the Balmer H$\alpha$ line to form, we turn to a geometrical argument. 

The area of the column along the line of sight which subtends the white dwarf is $\sim 2 \times H  \times R_{\text{wd}}$ and the volume is $2\times H \times \Delta R \times R_{\text{wd}}^2$, where $H$ is the gas height and $\Delta R = R_{\text{out}} - R_{\text{in}}$ is the gas extent in units of $R_{\text{wd}}$, with an outer radius, $R_{\text{out}}$, and an inner radius, $R_{\text{in}}$. The column density is equal to the number density times the volume of the gas column, divided by the area of the column along the line of sight which subtends the white dwarf. Solving for $n_{\text{H}}$, 
\begin{equation}
    n_{\text{H}} \lesssim \frac{N_{\text{H}}}{\Delta R \cdot R_{\text{wd}}}.
\end{equation}

We do not know the H number density, so we choose to explore a range such that $-1 \leq \log (n_{\text{H}}/\text{cm}^{-3}) \leq 10$. The minimum value is typical of the diffuse H II ($n_{\text{H}} \sim 0.3-10^4 \text{ cm}^{-3}$) and warm neutral medium ($n_{\text{H}} \sim 0.6 \text{ cm}^{-3}$) phases of the ISM \citep{Draine2011}. For a maximum plausible value, we rely on observations of specific circumstellar disks around protoplanetary and transition disks.\footnote{The maxmimum possible value of $N_{\text{H}}$ for TW Hydrae is $\log (N_{\text{H I}}/1 \text{ cm}^2) \sim 19.75$  \citep{Herczeg2004}, and the H column density for $\beta$ Pictoris is $\log (N_{\text{H}}/1 \text{ cm}^2) \sim 18.6\pm 0.1$, \citep{Wilson2017}.} For gas with a height of $10\%\,R_{\text{wd}}$, extending from $100\, R_{\text{wd}}$ to $200\, R_{\text{wd}}$, with $\log N_{\text{H}}\sim19$, $n_{\text{H}} \lesssim 10^9 \cm^{-3}$. We therefore explore the dependence of our models on the hydrogen density with a grid of models that have $-1 \leq \log (n_{\text{H}}/\text{cm}^{-3}) \leq 10$. For each model, the hydrogen density is constant with distance from the star. The most likely model is the one that minimizes the amount of H while allowing for metal lines to form.

\subsubsection{C-S Gas Abundances}

Calcium is the only C-S gas with a positive detection in WD 1124-293. Therefore, we consider the abundances of elements relative to Ca. Table \ref{tab:abund} lists the elements that are typical polluters of WDs for which we have photospheric abundance limits, and are thus explored with our modeling. We focus on the strongest optical transitions for these species that have photospheric detection (Mg I 3838 and Fe II 3228), and detection upper limits (K I 4043, Ni I 3480, Mn I 4032, Al I 3961, Si I 3905, Na I 5890). All other elements from He to Zn are left at the default solar composition values, relative to Ca. We approach the abundances in this way for two reasons. First, the ratios of potential metals of interest are very similar for a solar and chondritic composition (see Figure \ref{fig:abuns}). Second, there is thus far only one detected C-S feature in the optical part of the spectrum for WD 1124-293. A UV spectrum would likely show more absorption features that could be used to better constrain the C-S  metal abundances (see Figure \ref{fig:uv}).

\begin{figure}
    \centering
    \includegraphics[scale=0.4]{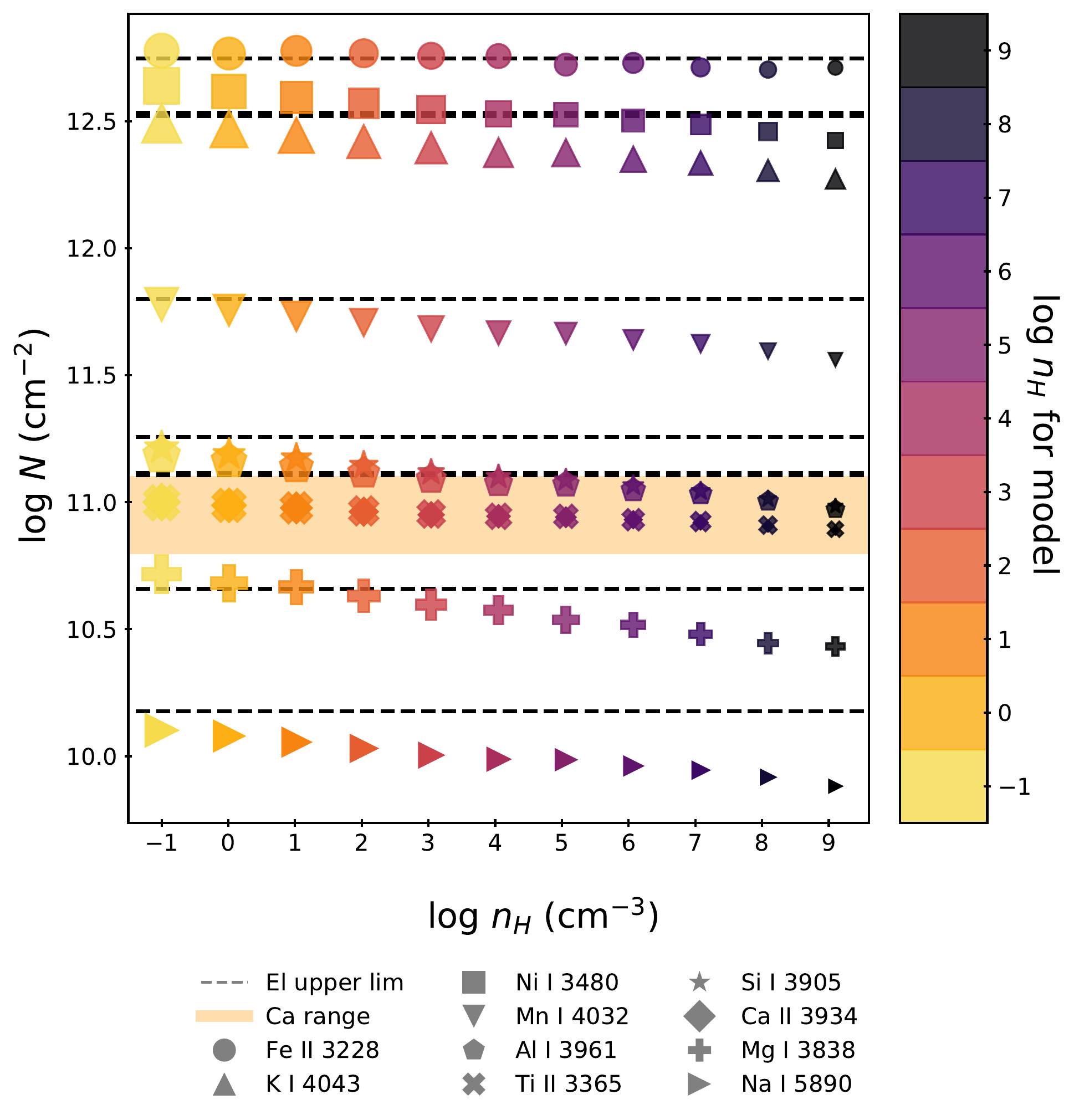}
    \caption{The column densities for a number of metal species in Cloudy models over hydrogen number densities from $\log n_H: -1 \text{ to } 9$ [cm$^{-3}$].  The orange horizontal bar highlights the range of expected Ca II column densities (\cite{Debes2012} and this work), and the dashed lines highlight the maximum column densities for other species. The size of each marker is related to the absolute value of the element abundance. The metal abundances are not sensitive to the number density hydrogen.}
    \label{fig:column}
\end{figure}

We use the C-S column density of Ca II 3934 to constrain the modeling by varying the abundance ratios of metals relative to calcium. Beginning with a hydrogen number density $n_{\text{H}} = 0.1 \cm^{-3}$ (a lower limit inspired by low density ISM regions), we first find abundance ratios that, when paired with the overall hydrogen density, result in optical depths that lead to a calculated absorption line with a depth at the $3\sigma$ limit to the observed continuum. For the rest of the metals considered, we ensured that their abundances were such that no lines were formed (see Figure \ref{fig:line}). We then fix those abundance ratios and vary the hydrogen density over 10 orders of magnitude, while increasing the abundance of Ca relative to H to explore the model dependence on the amount of hydrogen present. The resulting grid of models is presented in Figure \ref{fig:column}.

\subsection{Cloudy Model Output}

We choose to save the general overview, line optical depths, and line population for each Cloudy run. The general overview contains an output transmitted spectrum, but it does not include line broadening mechanisms, such as micro thermal motion, macro circular motion, or instrument effects, for an unresolved line. The net transmitted spectrum from Cloudy can however be used to check the validity of a set of input parameters (by investigating whether other absorption or emission features are produced), and to predict a spectrum ranging from the far UV to near IR (see Discussion). 

We use the optical depth and column density to compare Cloudy models to MIKE spectrum. We save the optical depths, $\tau$, for all species with $\tau>0.001$ to calculate an absorption line profile, and the populations of upper and lower levels for all lines to calculate the column densities of different species.  The column densities are calculated for each line by multiplying the lower level population per zone by the length of the zone. Zones are automatically calculated by Cloudy. We only consider the lower level populations in the column density calculation because we are not in a regime where stimulated emission is important.

% for figure above:
%The black vertical line  [coming] shows our upper limit for the H number density.

%%%%%%%%%%%%%%%%%%%%%%%%%%%%%%%%%%%%%%%%%
%       
%      R E S U L T S     +      D I S C U S S I O N
%
%%%%%%%%%%%%%%%%%%%%%%%%%%%%%%%%%%%%%%%

\section{Results and Discussion}\label{sec:disc}

We investigate the gas toward WD 1124-293 to explore the conditions necessary to produce the observed absorption. We further constrain the location and amount of Ca present in the C-S gas, reproduce the observed C-S Ca K absorption line profile and place upper limits to the amounts of H, Mg, Ca, Fe, and other metals present around WD 1124-293 by using Cloudy to determine their optical depths and expected line column densities. We place limits on the total amount of gas and show the temperature profile through the model disk. We then connect these results to other similar studies and argue for future UV observations. %present our main results and discuss their implications. 

%  H I R E S     A N D     S U B L I M A T I O N     R A D I U S

\subsection{Location of the gas}
From the new HIRES observations, we show that the previously detected C-S and photospheric Ca gas is still present. Calcium in the atmosphere of WD 1124-293 has a settling time of $\sim\!1000$ years \citep{Dufour2017}, so it is expected for the photospheric line to persist.

With the new observations we also find that the gas is farther away from WD 1124-293 than previously detected \citep{Debes2012}. Gaseous disks can accrete inward and spread outward due to angular momentum transport by turbulent viscosity \citep{Metzger2012}. However, why do we not observe gas at radii less than 100 $R_{\text{wd}}$ if the gas is created through sublimation at smaller radii? Even with no IR excess due to dust grains, if the particle density is low enough, it is possible that destructive grain-grain collisions near the tidal disruption radius could be creating gas \citep{Jura+2007} at $R_{\text{Kep}}\sim 100 R_{\text{wd}}$. Alternatively, there could be a collection of sub-micron grains, which are inefficient emitters, and would thus sublimate at much larger distances than what would be assumed in a blackbody approximation. Emission features due to Ca in the spectra of SDSS J$1228+1040$ show that the gas in this system is emitting at a range of radii ($0.6 - 1.2R_\odot$ or $50 - 100R_{\text{wd}}$), with a concentration at $\sim100\,R_{\text{wd}}$ \citep{Manser2016}.

\subsection{Column Densities} 

We determine the column density of the C-S Ca gas by finding the range of optical depths that lead to an absorption feature at a depth of $\pm3\sigma$ relative to the depth of the Ca C-S line in the MIKE dataset, resulting in $\log N(\text{Ca}) = 11.01^{+0.09}_{-0.26}$ (\citealt{Debes2012} find $\log N(\text{Ca}) = 11^{+0.1}_{-0.2}$). The downward trend in column density with increasing hydrogen number density seen in Figure \ref{fig:column} is due to a decrease in the calculated electron density. To investigate the dependence of our modeling on hydrogen, we explored 10 orders of magnitude in hydrogen number density, fixing the relative abundance ratios for all elements. The hydrogen number density is degenerate with the hydrogen abundance, so we are only able to constrain the amount of hydrogen with a geometrical argument described in Section 4.2.2. The upper limits to the abundances of these different species are presented in Table \ref{tab:abund}. 
\subsection{Line Profiles}

The Ca C-S absorption line is only marginally resolved, so we construct the absorption line profile by taking the convolution of (1) a Voigt line profile broadened due to Maxwellian distributed velocities and a gas temperature determined by Cloudy and (2) a Gaussian kernel with a width determined by the resolution of the HIRES spectrum.\footnote{If the absorption line were resolved, we would also convolve the Voigt line profile with a velocity profile, $\phi(v)$ describing the bulk motion of the gas disk.} The Voigt line profile is approximated using a series expansion as $\phi(x)\propto \exp(-x^2) + a/(\pi^{1/2}x^2) + 2a/(\pi^{3/4}x^4)$, where $a$ is the dampening constant for the transition in question, $x =  (\nu - \nu_0)/\Delta \nu_{\text{Dopp}}$, $\nu_0$ is the center frequency for line, and $\Delta \nu_{\text{Dopp}}$ is the FWHM of the line. The output intensity is $I = I_{\text{in}} \exp[-\tau\phi(x)]$, where $\tau$ is the optical depth at line center. The output intensity considering the velocity profile is $\propto \exp(-\tau[\phi(x)*\phi(v)])$. 

\begin{figure*}
    \centering
    \includegraphics[scale=.95]{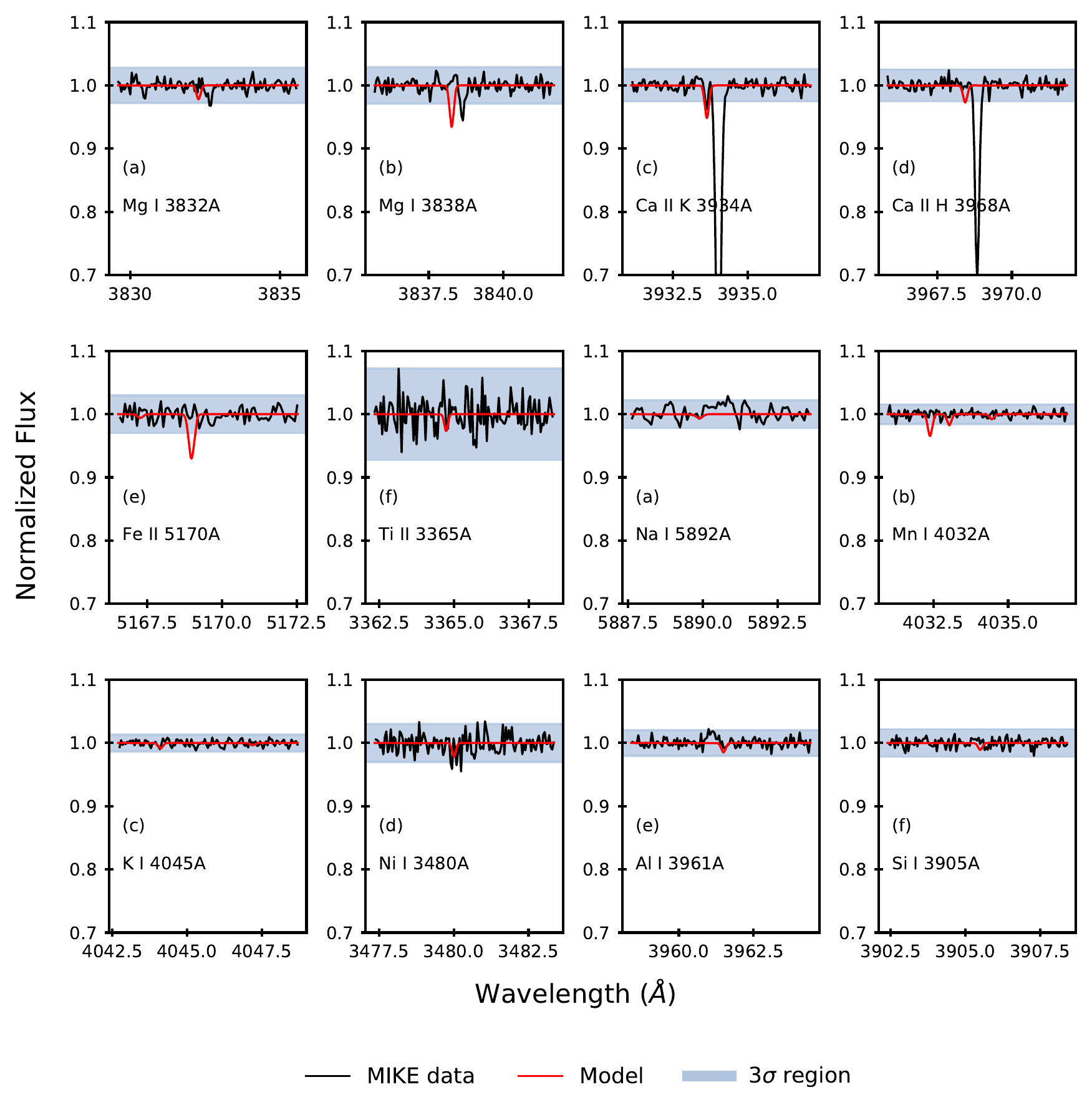}
    \caption{Calculated line profiles for species of interest. The calculated or model line profiles are shown in red. The black curves show the co-added MIKE spectra. The blue region highlights a $\pm3\sigma$ region for the MIKE spectra. The strengths of the model absorption lines are limited by column density upper limits estimated from the MIKE spectrum. For the Ca features, the deeper photospheric component is gravitationally red-shifted.}
    \label{fig:line}
\end{figure*}

We show the calculated line profiles for the strongest optical transitions of these metals in Figure \ref{fig:line}.

\begin{deluxetable}{lcccc}
\tablecaption{Photospheric and Circumstellar Metal Abundances and Masses \label{tab:abund}}
\tablecolumns{5}
\tablenum{4}
\tablewidth{0pt}
\tablehead{
\colhead{El} & \colhead{Phot } & \colhead{$\log \dot{M}(El)$} & \colhead{C-S} & \colhead{$\log M_{\text{C-S}}$} \\
\colhead{} & \colhead{} & \colhead{ $(\text{g yr}^{-1})$} & \colhead{} & \colhead{(g)}}
\startdata
%13.61621792, 14.29839347, 13.24377198, 14.53817227, 14.19785571,
%       13.33360388, 13.85773405, 14.01057069, 14.68768214, 14.10028415
H	& 1			    & -- & 1			& $<5.95$\\	%& 1 		& 14.857 \\
Na 	& $<-8.30$		& $<13.62$  & $<3.35$	& $<10.67$ \\
Mg	& $-7.689\pm0.024$	& $14.30$ & $<7.85$	& $<15.19$ \\
Al	& $<-8.80$	    & $<13.24$ & $<7.05$ 	& $<14.44$ \\
Si 	& $<-7.50$ 		& $<14.54$ & $<6.65$ 	& $<14.06$ \\
K	& $<-8.00$		& $<14.20$ & $<7.05$ 	& $<14.61$ \\
Ca	& $-8.872\pm0.188$	& $13.33$ & $1.357^{+0.086}_{-0.257}$ 	& $8.92^{+0.1}_{-0.3}$ \\
Ti	& $<-8.50$		& $<13.86$ & $<1.2$	& $<8.84$ \\
Mn	& $<-8.50$		& $<14.01$ & $<4.65$	& $<12.34$ \\
Fe	& $-7.814\pm0.130$	& $14.69$ & $<5.7$	& $<13.40$ \\
Ni	& $-8.40$ 		 	& $<14.10$ & $<5.35$	& $<13.08$ \\
\hline
%Total	&			&		&		& $<15.8$
Total	&			&		&		& $<15.38$
\enddata
%\tablenotetext{a}{Upper limits.}
\tablecomments{ Photospheric and circumstellar (C-S) abundances by number relative to hydrogen, $\log n(\text{El})/n(\text{H})$, for our model. We show the C-S abundances that corresponding to a model with $n_{\text{H}} = 0.1\cm^{-3}$.}
\end{deluxetable}

% H Y D R O G E N 

%\subsection{Hydrogen in a model that should have none}

%To investigate the dependence of our modeling on hydrogen, we explored 10 orders of magnitude in hydrogen number density, fixing the relative abundance ratios for all elements. The hydrogen number density is degenerate with the hydrogen abundance, so we are only able to constrain the amount of hydrogen with a geometrical argument described in Section 4.2.2. 

% O T H E R   M E T A L S  +  B E S T  F I T 
\begin{figure}
    \centering
      \includegraphics[scale=0.58]{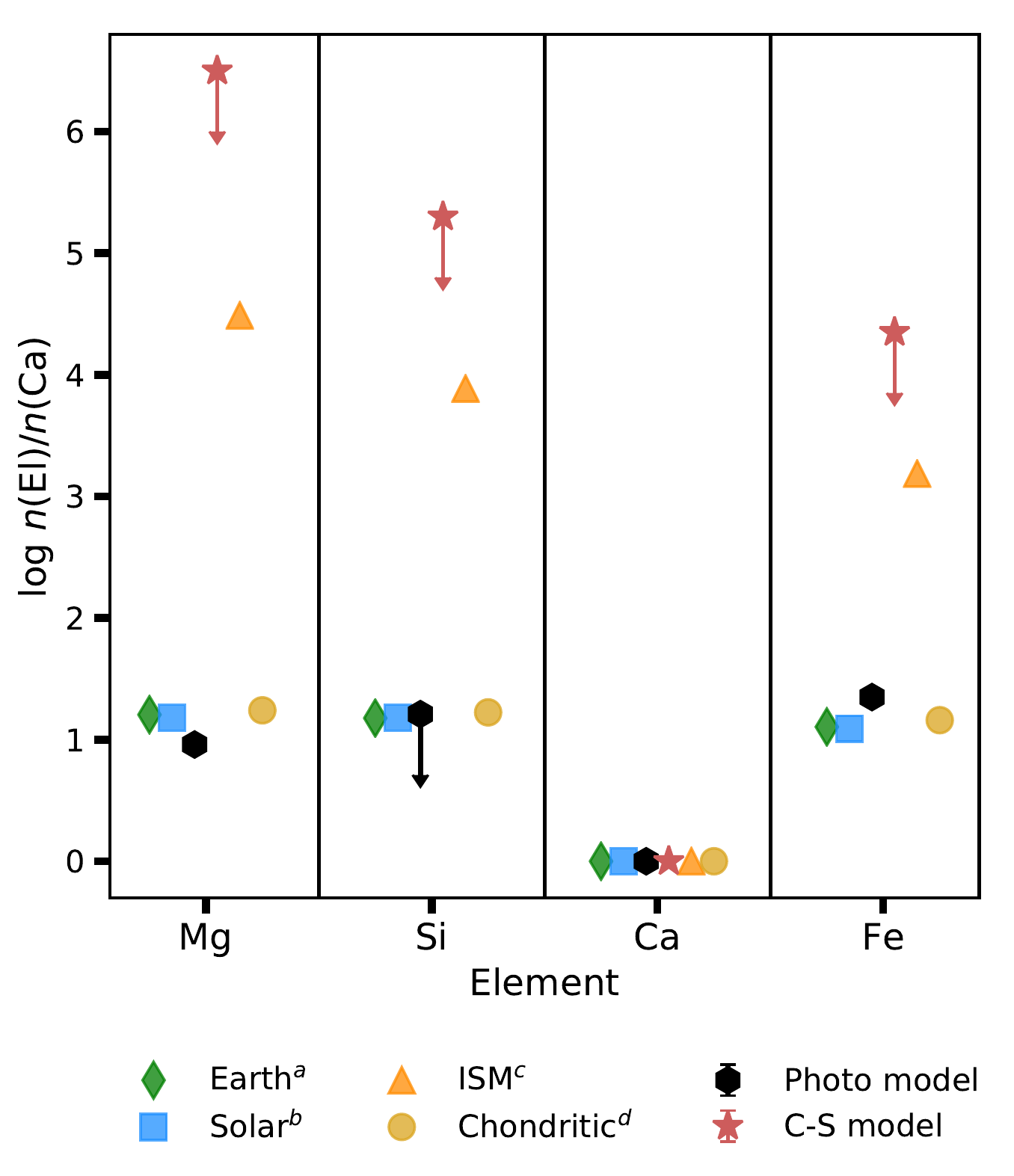}
    \caption{Abundances relative to calcium for metals typically present in photospheric pollution of WDs. We show the upper limits for our Cloudy model with $n_{\text{H}} = 0.1\cm^{-3}$ with the star markers and the abundances derived from the photospheric model considering diffusion with the black hexagons. We only had a marginal detection of silicon in the photosphere and provide an upper limit there. $^a$\cite{Allegre2001}, $^b$\cite{Holweger2001,Grevesse1998}, $^c$\cite{Cowie1986}, and $^d$\cite{Lodders2003}.}
    \label{fig:abuns}
\end{figure}

\subsection{Metals in the gas around WD 1124-293}

%For each n_H, we calculated a model with enough Ca to match the observed feature. For the rest of the metals listed in Table 4, we calculated upper limits to the abundances by increasing their abundance relative to Ca until a line was formed at <3-sigma below the normalized continuum. For the rest of the metals considered, we ensured that their abundances were such that no lines were formed (See Figure 5)

%(how is this different than the sentence before? IF they were just fixed to “solar” relative to Ca, you can just say that).

%Our most likely, upper limit model has enough Ca to match the observed feature, just enough of the metals in Table \ref{tab:abund} to produce an absorption feature at a depth of 3$\sigma$ from the normalized continuum, and avoids line formation for all other species (see Figure \ref{fig:line}). 
%For each $n_H$, we calculated a model with enough Ca to match the observed feature. For the rest of the metals listed in Table \ref{tab:abund}, we calculated upper limits to the abundances by increasing their abundance relative to Ca until a line was formed at $3\sigma$ below the normalized continuum. For the rest of the metals considered, we ensured that their abundances were such that no lines were formed (see Figure \ref{fig:line}). 
The calculated calcium K line column density corresponds to an  abundance of 1.35 relative to hydrogen when $n_{\text{H}} = 0.1\cm^{-3}$. As we increased the hydrogen number density, we let the calcium and hydrogen abundances vary from $-8.90 < \log n(\text{Ca})/n(\text{H}) < 1.35$ and $-1 < \log n(\text{H}) < 11$. For the lowest H number density, the abundances of Mg and Fe that produce an absorption feature at a depth of 3$\sigma$ are $\log n(\text{Mg})/n(\text{H}) = 7.85$ and $\log n(\text{Fe})/n(\text{H}) = 5.70$. We then fix the ratios of Mg and Fe relative to Ca with increasing density, such that $\log n(\text{Mg})/n(\text{Ca}) = 6.5$ and $\log n(\text{Fe})/n(\text{Ca}) = 4.4$ for all models shown in Figure \ref{fig:column}. All other element abundances, $El$, are set such that $\log n(El)/n(\text{Ca})$ is constant for models with increasing $n_{\hi}$. Figure \ref{fig:column} shows how the column density for the strongest observable optical transition in a species varies with increasing H number density and fixed element abundance ratios.

\subsection{The mass of the gas around WD1124-293}
Given our geometrical constraints (inner radius, $R_{\text{in}}$ and outer radius, $R_{\text{out}}$) and assumed disk height, $H$, the volume of the gas disk is $V = \pi H ( R_{\text{out}}^2 - R_{\text{in}}^2 ) $, and the total gas mass $M_{\text{tot}}$ is given by 
\begin{eqnarray}
M_{\text{tot}} = V\cdot \left[ \sum \left(10^{n_{\text{H}}} \cdot 10^{\text{abn}} \cdot m_{\text{El}} \right)  +  10^{n_{\text{H}}} \right] \cdot m_{\hi},
\end{eqnarray}
where $n_{\text{H}}$ is the hydrogen number density, abn is the abundance relative to hydrogen, $m_{\text{El}}$ is the mass of an element in atomic mass units, and $m_{\hi}$ is the mass of the hydrogen atom in g.  Using the abundances and hydrogen density from our model with a maximum amount of hydrogen ($n_{\text{H}}=10^9\cm^{-3}$), we place an upper limit on the total gas mass, $\log M_{\text{tot}} \approx 16.12$ g or $\sim\!30$ times the mass of C-type asteroid 162173 Ryugu ($4.50\times10^{14}$ g, \citealt{Watanabe2019}). Using our model with a minimum amount of hydrogen ($n_{\text{H}}=10^{-1}\cm^{-3}$), the upper limit on the total gas mass is $\log M_{\text{tot}} = 15.38$ g or $\sim\!5$ times the mass of Ryugu. The lower limit is set by the total amount of mass required to produce the Ca feature, so the C-S gas mass, $\log M_{\text{C-S}}$ is constrained to $8.92 < \log M_{\text{C-S}} < 16.12$. Figure \ref{fig:abuns} shows the relative abundances of Mg, Si, Ca, and Fe for our best fit and the relative abundances of those same metals in the photosphere.

\begin{figure}
    \centering
    \includegraphics[scale=0.4]{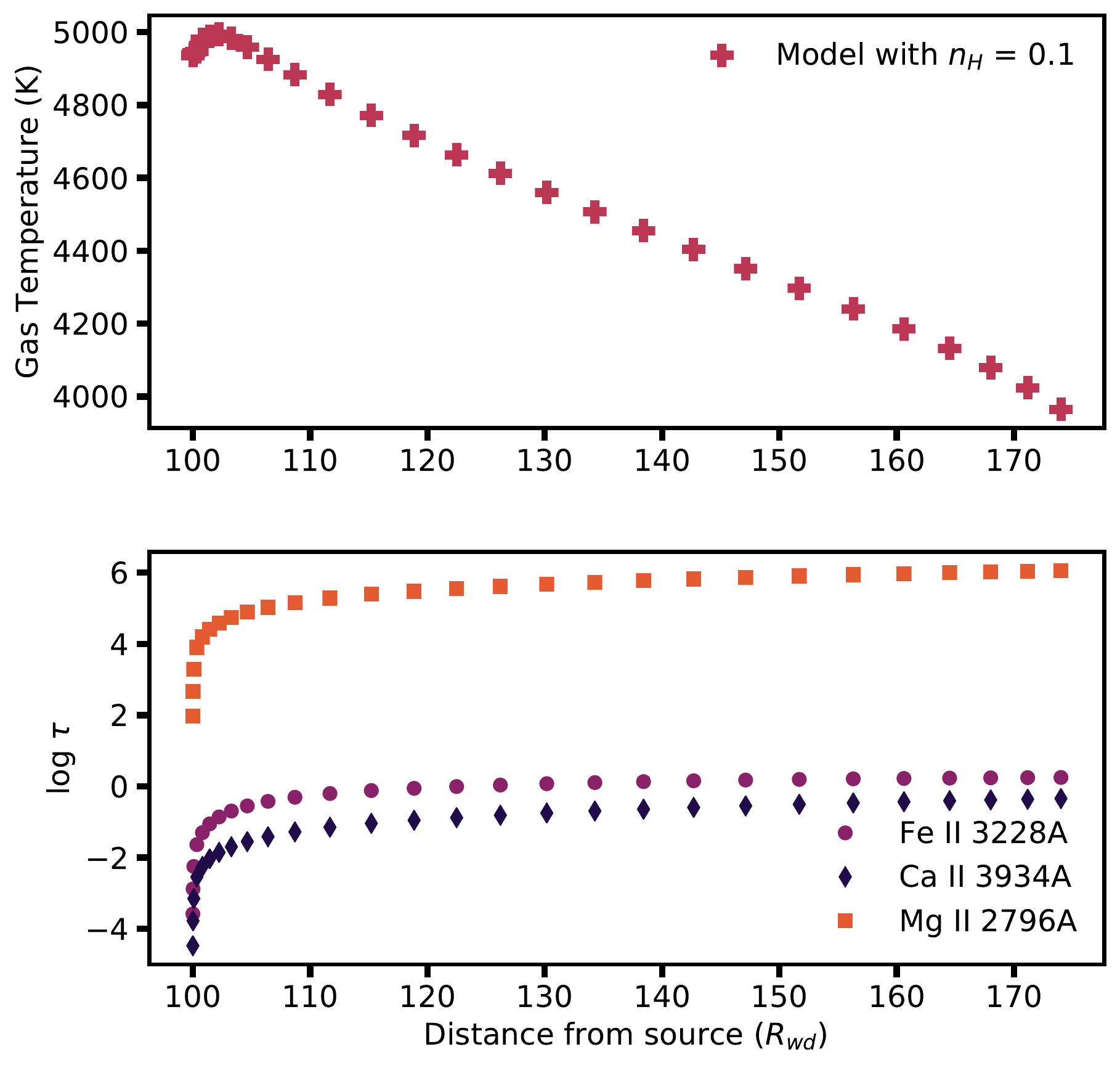}
    \caption{Top panel: The temperature profile through the disk. Some useful output from Cloudy is the ability to know the temperature profile within the cloud. Modeling of regions usually assume an isothermal environment which is not the case for many of the models. Lower panel: Optical depth with distance for the Ca K line, as well as the Fe and Mg lines responsible for heating and cooling, respectively. 
}
    \label{fig:temp}
\end{figure}

\subsection{Gas Temperature}

Cloudy also provides the temperature throughout the gas. For this model, the gas temperature ranges from $\sim\!4994$ K to $\sim\!3965$ K. We show the temperature profile for the best fit model in Figure \ref{fig:temp}, also including the optical depths of the two strongest lines at the same depths in the disk. 

In our Cloudy models, the heating is dominated by Fe II and the cooling is dominated by Mg II (see bottom panel of Figure~\ref{fig:temp}). The temperature of the gas is determined by Cloudy self-consistently and depends on heating and cooling mechanisms. Other efforts to model C-S around WDs tend to assume the temperature is isothermal, though at least one group calculates the temperature at different locations in the disk assuming a Shakura and Sunyaev viscous $\alpha$-disk using the accretion-disk code AcDc \citep{Hartmann2016,Nagel2004}. For a WD with $T_{\text{eff}} = 20\,900$ K, \cite{Hartmann2016} find the gas temperature decreases to a minimum value $\sim6000$ K a third of the way through the disk before rising to a maximum of $\sim6600$ K at the outer edge. The temperature profile of this gas disk is very different from that of WD 1124-293, as shown in the top panel of Figure~\ref{fig:temp}. For WD 1124-293, the gas temperature profile declines roughly as the $T\propto r^{-0.35}$, with a mild inversion at the inner edge of the disk. Further investigation is needed to understand this discrepancy. 
Additionally, for hotter WDs, the temperature range probed by Cloudy could be much larger than the 20\% changes seen for WD 1124-293. 

\begin{figure}
    \centering
      \includegraphics[scale=0.41]{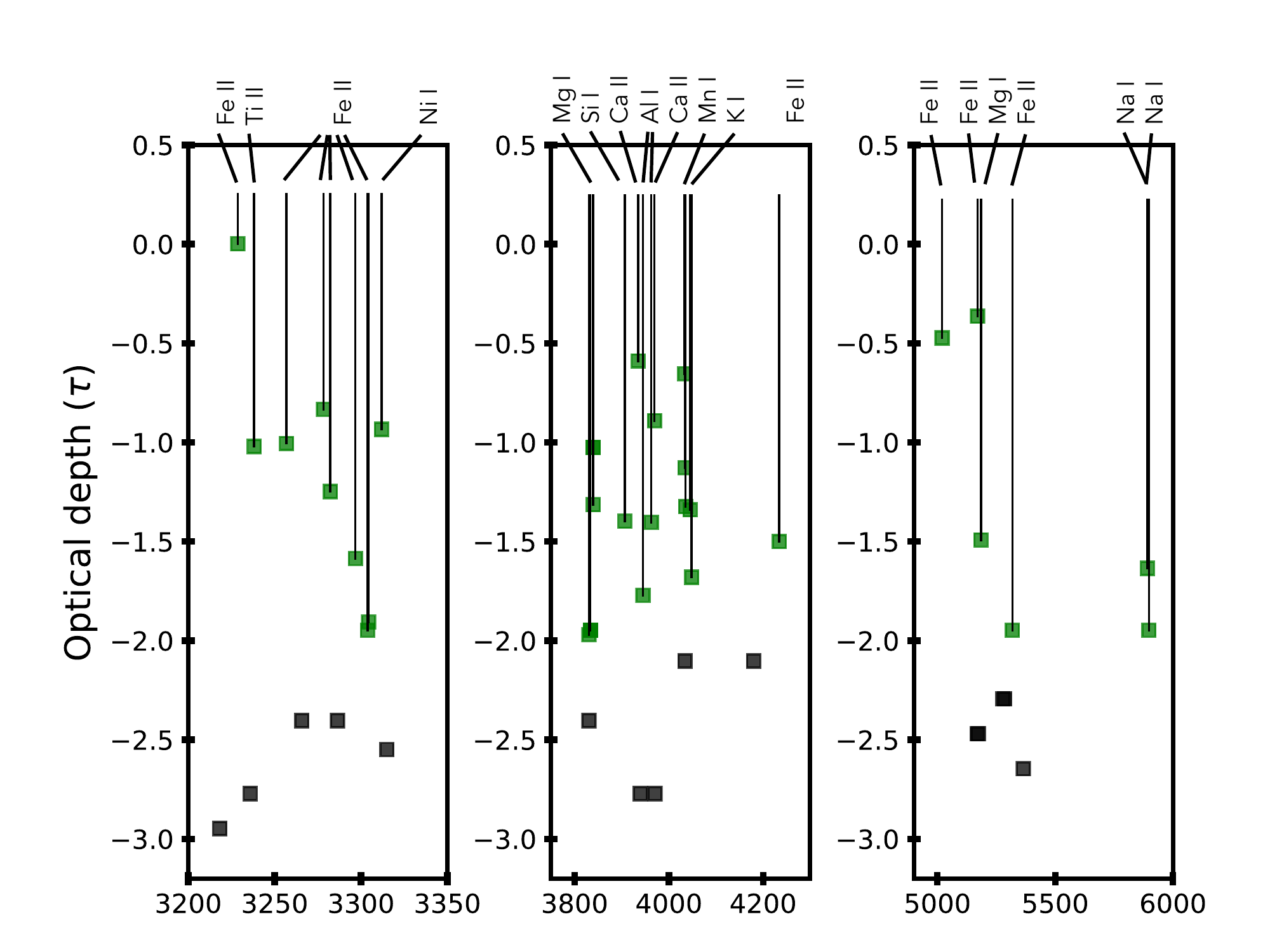}
    \caption{Optical depths versus wavelengths of the strongest species for our model with the maximum amount of hydrogen. The green squares highlight lines with optical depths $> -2.0$ and the black squares arise from different Mg, Ca, and Fe species. There are many Mg I lines at $\sim\!3800$\AA, though we only show one label for reading legibility.}
    \label{fig:taus}
\end{figure}

\begin{figure*}
    \centering
    \includegraphics[scale=0.66]{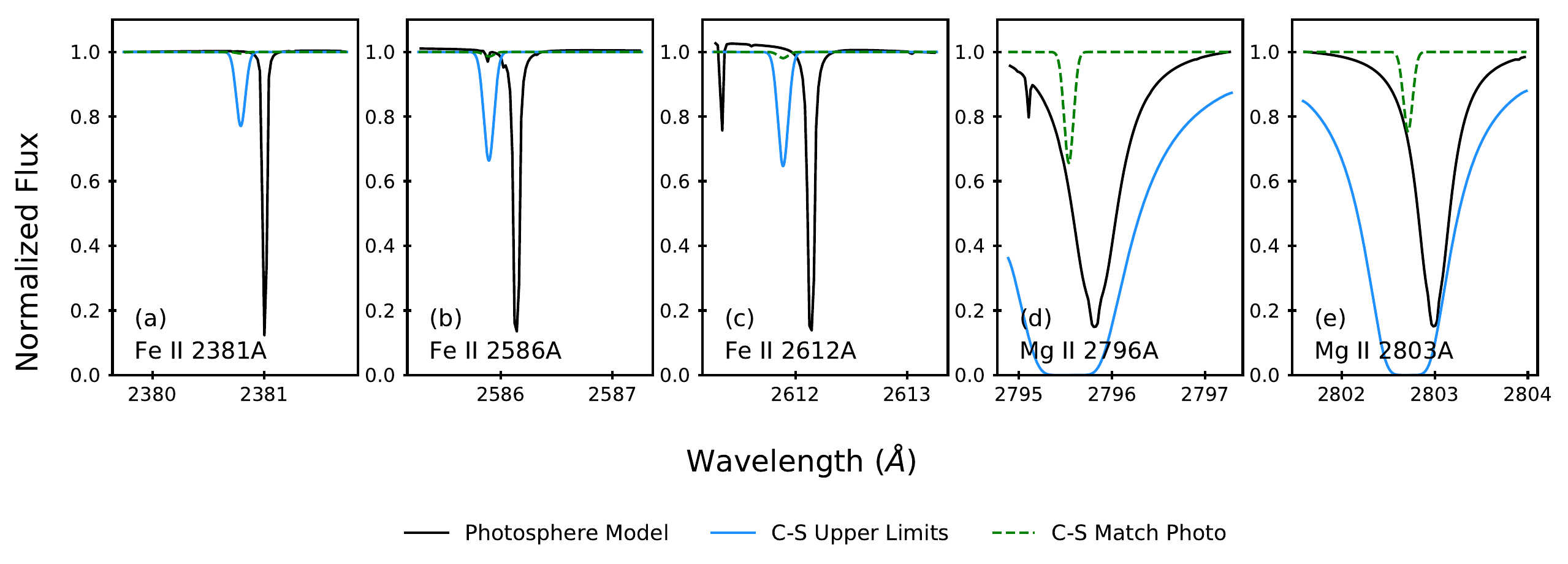}
    \caption{Predicted UV line profiles for Mg and Fe. The models presented in our results section extend into the UV, so we explore two cases.  In the first case, we set the abundance ratios of Mg/Ca and Fe/Ca to match the photospheric values. The solid black line is the best fit polluted photospheric model for WD 1124-293.  In the second case, we set the abundance ratios at the modeled upper limits. The expected profiles for Fe/Ca and Mg/Ca equal to the photospheric abundance ratios are shown with the dashed green line. The expected profiles for the C-S abundance upper limits are shown with a solid blue line. 
    }
    \label{fig:uv}
\end{figure*}

\subsection{Strongest transitions and a need for UV observations}

%Most asteroids are primarily made of O, Fe, and Si in addition to Mg and Ca. However, we note that 

The strongest expected lines in the optical come from the enhanced Mg, Ca, and Fe in our solar-like composition around WD1123-293 (see Figure \ref{fig:taus}). To obtain the abundances of heavy elements for the photosphere, we proceed in a similar way as described in \cite{Xu2019}, by computing grids of synthetic spectra for each element of interest using a pure DA atmospheric structure assuming the stellar parameters previously determined by fitting the photometric data and parallax measurement. The abundances are then obtained by minimizing $\chi^2$ between the normalized spectroscopic data and the grid spectrum. Trace amounts of metals have a negligible effect on the thermodynamic structure for the grid spectrum.

%Obtaining UV detections of Si, C, and O in particular can constrain how water-rich or C-rich these planetesimals are.

The strongest transitions for most of these species that would help constrain the gas composition and therefore, the sublimation temperature (O, Fe, Si, and Mg), are in the UV. Our polluted photospheric model for WD 1124-293 predicts very strong Mg and Fe absorption lines at UV wavelengths, and as such, our upper limits are likely too large (see Figure \ref{fig:uv}). 

FUV observations of other dusty WDs with $T_{\text{eff}}\geq12000$ K have shown absorption lines from as many as 19 unique elements (GD 362, \citealt{Xu2013}), providing detailed information about planetary material that orbits another star. For example, we can look for correlations between progenitor mass and elemental abundance, or seek to find correlations with specific elemental enhancements, such as Ca. With UV observations, better constraints would be placed on Mg and Fe, helping to further constrain the relative abundances of these metals.

% B O R I S :  What do we think about this system in the context of modeling polluted WDs. 

\subsection{Other work modeling WDs with Cloudy} 

\cite{Boris2019} use Cloudy to model the C-S gas of WDJ0914+1914 viewed in emission, which contains signatures of the disruption and subsequent accretion of a giant planet. WDJ0914+1914 has $T_{\text{eff}} = 27 743 \pm 310$K and its photosphere shows evidence of ongoing accretion of oxygen and sulfur. At this temperature, the C-S  gas is photoionized, and contains enough hydrogen for oxygen, sulfur, and H$\alpha$ emission lines to form. The emission features are doubly peaked, indicating the gas is in disk undergoing Keplerian rotation. In their work, the C-S  abundances for O and S are consistently determined by two independent methods for the first time. 

With this work, we follow up with the second instance to date of determining gas composition with two independent measurements, and show for the first time how to place constraints on abundances when only absorption features are present. This proof of concept for modeling absorption with Cloudy will be useful when applied to more complicated systems, such as WD 1145+017. 
% For Mg, Ca, and Fe, \cite{Boris2019} find abundance ratio upper limits of log (Mg/Ca) = 4.08 and log (Fe/Ca) = 2.15. --Siyi didn't like this. 

%For all of our models, the hydrogen density is constant. Due to the lack of an infrared excess \citep{Barber2016}, we exclude grains. The downward trend in column density with increasing hydrogen number density seen in Figure \ref{fig:column} is due to a decrease in the electron density.

% M A U D E :  Bring up Maude's paper (should be a paragraph discussing wd1145) 

\subsection{Our proof of concept in the context of WD 1145+017}
WD 1145+017 was first shown to have a transiting, disintegrating planetesimal by \cite{Vanderburg2015} and has been since been the subject of much observation (e.g. \citealt{Xu2016,Cauley2018,Croll2017,Xu2019,FortinArchambault}) due to days to week transit and spectroscopic variability. \cite{FortinArchambault} present a new characterization of WD 1145+017's absorption features, showing how a simple model of nested concentric rings with high eccentricity can help account for the majority of the observed features, which include asymmetrical line profiles and Doppler shifting of the absorption features. These features are also seen in a number of other polluted WDs showing emission lines (e.g., \citealt{Wilson2014, Manser2019, Boris2019}). 

The addition of high-resolution spectra to the analysis in \cite{FortinArchambault} helped disentangle closely spaced features that led to an overestimated prior abundance calculation in \cite{Xu2016}. A code like Cloudy could be useful in identifying components of such blends to obtain accurate abundances.  They also observe Si IV features in the UV at 1393.76\AA{} and 1402.77\AA{} and invoke an additional low eccentricity component to explain the presence of such highly ionized species. Cloudy would naturally be able to probe the conditions needed to produce Si IV features with its self-consistent microphysics.

\iffalse
The addition of high-resolution spectra to the analysis in \cite{FortinArchambault} helped disentangle closely spaced features that led to an overestimated prior abundance calculation in \cite{Xu2016}. 

\cite{FortinArchambault} show that metals in the C-S  gas and photosphere have roughly the same abundances and the chemical abundance pattern is consistent with the accretion of a rocky body with bulk Earth composition. They also observe Si IV features in the UV at 1393.76\AA{} and 1402.77\AA{} and invoke an additional low eccentricity component to explain the presence of such highly ionized species. A code like Cloudy could be useful in identifying components of such blends to obtain accurate abundances. 

 for a hotter WD, the temperature range probed could be much larger than the 20\% changes seen for 1124

Cloudy would naturally be able to probe how to generate Si IV with its self-consistent microphysics. 
\fi

 \section{Conclusion}\label{sec:conc}

With this work, we outline how to use the Cloudy radiative transfer code to model C-S  gas viewed in absorption around WD 1124-293. We create a Cloudy grid of models for C-S gas around WD 1124-293 to explore the abundances of elements from He to Zn relative to hydrogen, and obtain line optical depths, species column densities, and the temperature profile through the gas disk. Our best fit model minimizes the total amount of hydrogen, while still producing the observed Ca K C-S absorption feature. 

We detect photospheric absorption features due to Mg and Fe for the first time, determine a new location for the C-S gas, and place upper limits on abundances of other metals in the photosphere. The upper limit C-S abundance ratios of Mg, Si, and Fe to Ca are also consistent with the photospheric abundance ratios, and with a chondritic/bulk Earth composition.

With these models, we place constraints on the potential masses and abundances that could result in a spectrum dominated by calcium species for WD 1124-293, find that the C-S is likely not isothermal, and show that the Cloudy microphysics code, which is typically used to model active galactic nuclei and HII regions, can also be used to model C-S  gas absorption features of polluted white dwarfs. UV spectroscopic observations of WD 1124-293 are needed to further constrain the composition of its C-S gas. 

Looking forward, we intend to explore the properties of C-S gas around DA WDs with temperatures 6000 K to 27,000 K using Cloudy, and will the make the grid of gas properties publicly available (Steele et al. \textit{in prep}).

\acknowledgments
We thank the Referee for their helpful comments. We thank the W. M. Keck Observatory Visiting Scholars Program for hosting A. Steele for 7 weeks. Special thanks go to Dr. Sherry Yeh and Dr. Carlos Alvarez for their guidance during the program. Additional thanks go to Dr. Jordan Steckloff for helpful conversations about sublimation near white dwarfs; and to Dr. Drake Deming and Dr. Siyi Xu for providing financial support for this work.

This work has made use of data from the European Space Agency (ESA) mission
{\it Gaia} (\url{https://www.cosmos.esa.int/gaia}), processed by the {\it Gaia}
Data Processing and Analysis Consortium (DPAC,
\url{https://www.cosmos.esa.int/web/gaia/dpac/consortium}). Funding for the DPAC
has been provided by national institutions, in particular the institutions
participating in the {\it Gaia} Multilateral Agreement. This work is supported by the international Gemini Observatory, a program of NSF's NOIRLab, which is managed by the Association of Universities for Research in Astronomy (AURA) under a cooperative agreement with the National Science Foundation, on behalf of the Gemini partnership of Argentina, Brazil, Canada, Chile, the Republic of Korea, and the United States of America. Some of the data presented herein were obtained at the W. M. Keck Observatory, which is operated as a scientific partnership among the California Institute of Technology, the University of California and the National Aeronautics and Space Administration. The Observatory was made possible by the generous financial support of the W. M. Keck Foundation.

%% To help institutions obtain information on the effectiveness of their 
%% telescopes the AAS Journals has created a group of keywords for telescope 
%% facilities.
%
%% Following the acknowledgments section, use the following syntax and the
%% \facility{} or \facilities{} macros to list the keywords of facilities used 
%% in the research for the paper.  Each keyword is check against the master 
%% list during copy editing.  Individual instruments can be provided in 
%% parentheses, after the keyword, but they are not verified.

\facilities{Magellan(MIKE), Keck(HIRES)}

%% Similar to \facility{}, there is the optional \software command to allow 
%% authors a place to specify which programs were used during the creation of 
%% the manusscript. Authors should list each code and include either a
%% citation or url to the code inside ()s when available.

\software{astropy \citep{2013A&A...558A..33A},Cloudy \citep{Ferland2017}, PyAstronomy \citep{Czesla2019}
         }

\end{CJK}

\end{document}